\documentstyle[11pt]{article}
\title{\begin{flushright}
{\small NSF-ITP-93-132 \\
NUC-MINN-93/29-T \\
PITT-93-12 \\
October 1993 \\}
\end{flushright}
{\bf MESONS IN THE QUENCHED APPROXIMATION AT LOW TEMPERATURE}}
\author{{\bf A. I. Bochkarev}$^{\dagger}$ and
 {\bf J. I. Kapusta}$^{\ddagger}$ \\
  {\it Institute for Theoretical Physics} \\
  {\it University of California} \\
  {\it Santa Barbara, CA 93106}}

\date{}

\begin{document}

\maketitle

\begin{center}
Abstract\\
\end{center}

We use an effective low energy field theory to describe the coupling between
glueballs and mesons as an analog of the quenched approximation in lattice
QCD.  This allows us to study the temperature dependence of mesonic
screening masses at temperatures below the deconfinement phase transition.

\vspace{.5in}

\noindent $^{\dagger}$ On leave from: {\it Department of Physics and
Astronomy, University of Pittsburgh, Pittsburgh, PA 15260}\\
\noindent $^{\ddagger}$ On leave from: {\it School of Physics and
Astronomy, University of Minnesota, Minneapolis, MN 55455}

\vfill \eject

In the last decade there has been much interest and work on a possible
chiral/deconfinement phase transition in QCD at some temperature on the
order of 200 MeV \cite{conf,lattice}.  In the vicinity of this temperature
neither
perturbation theory with quarks and gluons nor effective field theories
with hadrons can give a reliable description of the thermodynamics.
Lattice QCD is the best quantitative tool available.  However, Monte
Carlo computations with dynamical quarks are extremely time consuming.
This has led to studies in the quenched approximation, which basically
means that operator averages are computed in an ensemble of gluonic fields.
Then Monte Carlo computations can proceed much more rapidly and with
much better statistics. Not only is the quenched approximation easy
to implement in the lattice simulations, it is also known to be a
rather good approximation at zero temperature {\it a posteriori}.
The masses of the
low-lying hadrons as computed with no dynamical quarks are very close
to the experimental values \cite{lattice}. This should be no surprise,
because many hadronic parameters are successfully calculated by means of QCD
sum rules \cite{shif}, which remain true within the quenched approximation. The
contribution of
nonvalence quarks is in the higher-order multi-loop diagrams of
the perturbative sector and is naturally small.  Nonvalence quarks
also are unimportant in many successful phenomenological theories like
the nonrelativistic quark model \cite{qmodel} and the MIT bag model
\cite{bmodel}. Motivated by the success of the quenched approximation
at zero temperature, one is naturally led to wonder about what kind of
matter this approximation corresponds to at finite temperature.

Consider, for example, the correlator of two vector currents
$J_{\nu}\,=\,\bar{\psi}\,\gamma_{\nu}\,\psi$ \cite{Shuryak}.
After integrating out the quark fields in full QCD one has
\begin{equation}
<{\cal T}\left[ J_{\mu}(x)\,J_{\nu}(0) \right]> \;\;=\;\;
\frac{ \int [dA] \;e^{{\cal S}(A)} \; {\rm Tr}\, \left\{ \gamma_{\mu}\,
D_{A}^{-1}\,\gamma_{\mu}\, D_{A}^{-1} \right\}
{\rm det}\left(D_{A}\right)}{\int [dA] \,
e^{{\cal S}(A)} \, {\rm det}\left(D_{A} \right)} \, ,
\end{equation}
where ${\cal T}$ stands for chronological ordering, ${\cal S}(A)$ is
the action of glufermion propagator in the external gauge field $A$.
The quenched approximation ignores the dependence of the quark
determinant on the gauge field and so the determinants in the numerator
and denominator cancel in this average.
This means that the only quarks which enter the
average are introduced via the current operators themselves; these may
be called valence quarks.  There are no additional internal quark loops.

As another example, one may compute the contribution of valence quarks
to the free energy of the system by
\begin{equation}
{\cal F}_{q} \;\;=\;\; T\; \frac{\int [dA] \;e^{{\cal S}(A)} \;
{\rm Tr} \, \ln \, (D_{A}^{-1}) }{\int [dA]
\, e^{{\cal S}(A)}} \, .
\end{equation}
The free energy ${\cal F}_{q}$ incorporates that part of the usual
loop expansion of the full QCD free energy which has only one
quark loop. The complete thermodynamic potential of QCD in the
quenched approximation is the sum of ${\cal F}_{q}$ and the free energy of
pure gluodynamics.

How can one describe the correlation function (1) phenomenologically
at low temperatures?
In full QCD low temperatures are dominated by pions since they are by
far the lightest hadrons.  In the quenched approximation there are no
dynamical pions, and so the thermodynamics will be dominated by the lowest
mass glueball.  Chiral symmetry no longer plays an important role \cite
{Leut}.  The pure glue theory has scale invariance which is broken by the
renormalization prescription.  Nevertheless, there is a memory of
the classical scale invariance which leads to an infinite set of Ward
identities.  These identities can be saturated by one field of scalar
gluonium, which can be thought of as a fluctuation of the
gluon condensate, in the same sense that the pion is the low energy
fluctuation of the quark condensate \cite{scale,Ellis}.  The effective
La\begin{eqnarray}
{\cal L}_{\chi}&=& \frac{1}{2}\left(\partial_{\mu}\chi\right)^2
-U(\chi) \, , \nonumber \\
U(\chi)&=&B\left[4\left(\frac{\chi}{\chi_0}\right)^4 \ln\left(
\frac{\chi}{\chi_0}\right) -\left(\frac{\chi}{\chi_0}\right)^4
+ 1 \right] \, .
\end{eqnarray}
This Lagrangian reproduces the scale anomaly \cite{anomaly}.  The
potential is minimized when $\chi$ = $\chi_0$, and $U(\chi_0)-U(0) = -B$, where
the vacuum energy $B$ is related to the gluon condensate by
$<\frac{\beta}{2g}F_{\mu\nu}^a F^{\mu\nu}_a> = -4B $.
Anticipating that the gluonic condensate will change with temperature T,
we write $\chi = \bar{\chi} + \phi$, where $\bar{\chi}$ is the T-dependent
condensate and $\phi$ is the fluctuation.  The $\phi^2$ term in the
Lagrangian gives the glueball mass as
\begin{equation}
m_G^2(\bar{\chi}) \, = \, \frac{16B}{\chi_0^2}\left(\frac{\bar{\chi}}
{\chi_0}\right)
^2 \left[3\ln\left(\frac{\bar{\chi}}{\chi_0}\right)
+ 1\right] \, .
\end{equation}
In the vacuum $\bar{\chi} = \chi_0$ and $m_G^2(\chi_0) = 16B/\chi_0^2$.

Now consider the coupling of a spin-1 meson $V$ to the glueball in
the low energy effective theory.  Since all the scale
breaking is realized by the glueball Lagrangian (3), the coupling
must be \cite{Ellis}
\begin{equation}
{\cal L}_V \,=\, -\frac{1}{4}V_{\mu\nu}V^{\mu\nu} + \frac{1}{2}
\lambda^2\chi^2 V_{\mu}V^{\mu} \, ,
\end{equation}
where $\lambda$ is a dimensionless coupling constant.  Writing the
glueball field as a condensate plus a fluctuation, the mass is
$m_V(\bar{\chi}) = \lambda\bar{\chi}$.  Each spin-1 boson has its
own coupling constant $\lambda$ which would correspond to the mass
as measured by quenched lattice QCD at zero temperature.
The correlator of two vector fields is
\begin{equation}
<{\cal T}\left[ V_{\mu}(x)\,V_{\nu}(0) \right]> \;\;=\;\;
\frac{\int [d\chi{\rm det}^{-1/2}\left({\cal D}_{\mu\nu}\right) \;
{\cal D}_{\mu\nu}^{-1}(\chi, x)} {\int [d\chi] \,
e^{{\cal S}(\chi)} \;{\rm det}^{-1/2}\left({\cal D}_{\mu\nu}\right)} \, ,
\end{equation}
where the propagator ${\cal D}_{\mu\nu}^{-1}(\chi, x)$ in
the external field $\chi$ is the inverse of the Lagrangian
differential operator in (5).  Quenched QCD corresponds to
ignoring the $\chi$-dependence of the determinant.
The expansion of ${\cal D}_{\mu\nu}^{-1}(\chi, x)$ in a power series
in $\chi$ generates the usual Feynman expansion of the propagator in
the Dirac representation with no internal loops of the $V$-field.

The free energy that describes hadronic matter in the quenched
approximation allows for a representation similar to (2).
It contains thermal loops of hadrons with hadronic propagators in the
external fields of glueballs.

At low temperature it suffices to compute the meson
self-energy to one-loop order; more loops either give contributions
which are higher order in the particle densities, which are unimportant,
or they renormalize zero temperature coupling constants and masses.
There are cubic and quartic interactions between fluctuations in the
glueball field $\phi$ and the meson field $V$.  The one-loop self-energy
diagrams are shown in fig. 1.  The tadpole involves nonvalence quarks
and so is not included in the quenched approximation.  In
Euclidean space the self-energy is
\begin{eqnarray}
\Pi_V^{\mu\nu}(k)&=&\delta^{\mu\nu}\lambda^2T\sum_{p_4}
\int \frac{d^3p}{(2\pi)^3}\frac{1}{p^2+m_G^2(\bar{\chi})} \nonumber \\
&-&4\lambda^2T\sum_{p_4}
\int \frac{d^3p}{(2\pi)^3}\frac{1}{(p-k)^2+m_G^2(\bar{\chi})}
\frac{\delta^{\mu\nu}m_V^2(\bar{\chi})+p^{\mu}p^{\nu}}
{p^2+m_V^2(\bar{\chi})} \, .
\end{eqnarray}
The finite temperature contribution to the static, infrared limit of
the time-time component, relevant for screenin\begin{equation}
\Pi_V^{44}(k_0=0,{\bf k}\rightarrow 0)\,=\,\lambda^2\int\frac{d^3p}
{(2\pi)^3} \left[ \frac{4{\bf p}^2}{m_G^2(\bar{\chi})-m_V^2(\bar{\chi})}
\left( \frac{n_V}{\omega_V}-\frac{n_G}{\omega_G}\right)
-3\frac{n_G}{\omega_G} \right] \, ,
\end{equation}
where the $n$ are Bose-Einstein distributions, $\omega_V = \sqrt{
{\bf p}^2 + m_V^2(\bar{\chi})}$, and similarly for the glueball.

The screening mass is
\begin{equation}
m_{V,\,{\rm screen}}^2\,=\,m_V^2(\chi_0)\left(\bar{\chi}/\chi_0\right)^2
+\Pi_V^{44}(k_0=0,{\bf k}\rightarrow 0) \, .
\end{equation}
Note that there are two contributions to the screening mass: the mean field
and the one-loop scattering.  The shift in the mean field at low temperature
can be computed to first order in the density of thermal glueballs to be
\begin{equation}
m_V^2(\chi_0)\left( \frac{\bar{\chi}}{\chi_0} \right)^2 \, = \,
m_V^2(\chi_0) - 5\lambda^2\int\frac{d^3p}{(2\pi)^3}
\frac{n_G}{\omega_G} \, .
\end{equation}
Thermal fluctuations act to reduce the magnitude of the condensate
as one would expect.  To lowest order it suffices to use the zero
temperature masses on the right side of eqs. (8) and (10).

To see how important finite temperature effects are we use
$B = (200\,{\rm MeV})^4$ and $m_G = 1\,{\rm GeV}$.  The
coupling $\lambda$ is chosen so as to give the appropriate zero temperature
meson mass.  Numerical results are shown in fig. 2 for the $\rho$
meson.  The shift in the mean field contribution at finite temperature
is comparable in magnitude to the one-loop scattering contribution.
The change in the mean field with temperature acts to decrease the
$\rho$ mass, whereas the scattering term can increase or decrease it
depending on the relative magnitude of $m_G$ and $m_{\rho}$ in the
vacuum.  Both contributions are
{\it very} small because the Boltzmann factor $\exp(-m/T)$ provides a
big suppreswe have plotted the mass up to $T = 200$ MeV, the calculation cannot
be trusted to quite such high temperatures.  Calculation of the
equation of state of pure glue SU(3) on the lattice results in a
first order phase transition at a critical temperature of $225\pm10$
MeV \cite{lattice}.  As this temperature is approached, higher mass
glueballs become
increasingly important.  The density of glueballs becomes large and
our low temperature/low density approximation breaks down.  Probably
there is an exponentially increasing glueball mass spectrum, and the
thermodynamics of such a system of interacting glueballs becomes
difficult to treat \cite{Hag}.

The same calculation can be done for spin-0 mesons.  The one-loop
self-energy diagrams are the same as for spin-1 mesons.  Eq. (8) gets
replaced by
\begin{equation}
\Pi_S(k_0=0,{\bf k}\rightarrow 0)\,=\,\lambda^2\int\frac{d^3p}
{(2\pi)^3} \left[ \frac{n_G}{\omega_G} -
\frac{4m_S^2}{m_G^2-m_S^2}
\left( \frac{n_S}{\omega_S}-\frac{n_G}{\omega_G}\right)
\right] \, .
\end{equation}
The shift in the screening mass for any spin-0 meson whose vacuum mass
lies in the mid-GeV range and above would be as small as for the $\rho$
meson.

In conclusion, we have discussed how the quenched approximation to
QCD at low temperatures can be modelled by a low energy effective
Lagrangian.  At temperatures well below $T_c$ the lightest scalar
glueball, which describes the breaking of scale invariance
in QCD, plays the dominant role.  In the quenched approximation it
is the gluonic condensate which gives mass to the usual hadrons at
zero temperature.  The glueball part of the Lagrangian has two
parameters which are determined at zero temperature.  Each meson has
its own coupling $\lambda$ to the glueball field, which is fixed
by computing the meson mass at zero temperaturThen the low temperature
dependence of all hadronic screening masses
are {\it predicted} from this low energy effective theory.  It is very
interesting that in the quenched approximation chiral symmetry, and
hence dynamical pions, play essentially no role.  On the contrary, previous
calculations of mesonic screening masses at finite temperature
using QCD sum rules \cite{sumT1,sumT2} or effective low energy Lagrangians
\cite{Leut,Minn} have shown
the dominance of pion dynamics in the full theory.  The quenched
approximation allows us to focus on the nature and dynamics of broken
scale invariance as opposed to chiral symmetry.
Unfortunately, finite temperature effects on mesonic screening masses
are rather small at low temperatures in the quenched approximation.
This is a consequence of the substantial difference between the
spontaneous breaking of scale invariance, which leads to a very heavy
scalar gluonium with mass of order 1 GeV, and the spontaneous breaking
of chiral symmetry, which leads to very light pions.
Although we have focussed on mesonic screening masses, there are many
other quantities that can be studied at low temperatures, such as the
equation of state and the screening masses of baryons.
We eagerly look forward to high statistics measurements
of quenched QCD on the lattice at finite temperature \cite{Bielefeld}.

\section*{Acknowledgments}

We gratefully acknowledge discussions with F. Karsch and B. Petersson
on lattice QCD in the quenched approximation.
This work was supported by the U.S. Department of Energy under grant
DOE/DE-FG02-87ER40328 and by the U.S. National
Science Foundation under grants PHY89-04035 and PHY90-24764.

\section*{Figure Captions}

\noindent Fig. 1.  One-loop contributions to the meson self-energy.
The tadpole diagram is not included in the quenched approximation
because it involves nonvalence quarks.\\

\noindent Fig. 2.  Change in the $\rho$ meson screening mass with
temperature in the quenched approximation.  The dashed line is
the mean field contribution.  The solid line includes both the
mean field and the one-loop scattering contributions.

\end{document}